1# Detection of Acetone as a Gas Biomarker for Diabetes Based on Gas Sensor Technology

Jiaming Wei, Tong Liu, Jipeng Huang, Xiaowei Li, Yurui Qi, Gangyin Luo

*Abstract*—With the continuous development and improvement of medical services, there is a growing demand for improving diabetes diagnosis. Exhaled breath analysis, characterized by its speed, convenience, and non-invasive nature, is leading the trend in diagnostic development. Studies have shown that the acetone levels in the breath of diabetes patients are higher than normal, making acetone a basis for diabetes breath analysis. This provides a more readily accepted method for early diabetes prevention and monitoring. Addressing issues such as the invasive nature, disease transmission risks, and complexity of diabetes testing, this study aims to design a diabetes gas biomarker acetone detection system centered around a sensor array using gas sensors and pattern recognition algorithms. The research covers sensor selection, sensor preparation, circuit design, data acquisition and processing, and detection model establishment to accurately identify acetone. Titanium dioxide was chosen as the nano gas-sensitive material to prepare the acetone gas sensor, with data collection conducted using STM32. Filtering was applied to process the raw sensor data, followed by feature extraction using principal component analysis. A recognition model based on support vector machine algorithm was used for qualitative identification of gas samples, while a recognition model based on backpropagation neural network was employed for quantitative detection of gas sample concentrations. Experimental results demonstrated recognition accuracies of 96% and 97.5% for acetone-ethanol and acetone-methanol mixed gases, and 90% for ternary acetone, ethanol, and methanol mixed gases.

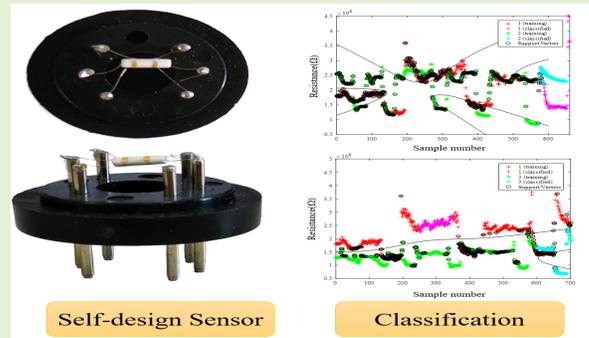

## I. Introduction

DIABETES is a chronic disease [1]. The number of patients worldwide remains high and continues to grow. Improving the diagnosis of diabetes is urgent once cases are counted and intuitive data is generated. Compared to venous blood testing which carries the risk of disease transmission and urine testing which is cumbersome and not applicable to all patients, breath testing offers the advantages of being rapid, convenient, and non-invasive.

From the emergence of diabetes to the continuous improvement of diagnostic technologies, doctors have conducted multifaceted research and discovered that the odorous gases in exhaled breath actually consist of Volatile Organic Compounds (VOCs). These compounds are noticeable because the onset of the disease affects normal metabolism in the body, causing these compounds to be exhaled by patients and their relative levels to increase. These compounds are associated with certain diseases in the body, and by detecting and analyzing their levels, patients can undergo preliminary screening. This testing method also needs to clarify the correlation between these compounds and the diseases they correspond to [2]. The World Health Organization collectively defines these compounds associated with diseases as "biomarkers." Some biomarkers and their related diseases are shown in Table I.

TABLE I
BIOMARKERS AND RELATED

| Biomarkers | Disease |
| --- | --- |
| Acetone | Diabetes, lung cancer, reduced dietary fat |
| Carbon monoxide | Oxidative stress, respiratory infections, anemia |
| Carbon disulfide | Schizophrenia, coronary artery disease, arterial disease |
| Methyl mercaptan | Bad breath |
| Nitric oxide | Asthma, bronchiectasis, hypertension, rhinitis |
| Sulfur compounds | Liver disease, malaria, cancer |
| Ethanol | The production of intestinal bacteria |
| Methanol | Neurological disorders |

(Corresponding author: Jipeng Huang.)
Jiaming Wei, Tong Liu, Jipeng Huang, Xiaowei Li and Yurui Qi are with the School of Physics, Northeast Normal University, Changchun 130024, China (e-mail: weijm713@nenu.edu.cn ).
Gangyin Luo is with the Suzhou Institute of Biomedical Engineering and Technology, Chinese Academy of Sciences, Suzhou 21563, China (e-mail: luogy@sibet.ac.cn).

The detection equipment requirements for acetone gas in this study indicate that gas sensor-based detection technology has significant technical advantages. With technological advancements, machine learning algorithms have been continuously developed, applied in multiple scenarios, improved horizontally and vertically, and combined with chemical sensor detection technology to provide the advantages of convenience, speed, and intelligence [3], [4], [5], [6]. This combination makes it possible to design and develop cost-effective, miniaturized, portable non-invasive breath testing systems. The significance of this study can be elaborated on from the following two aspects.

From a theoretical contribution perspective: by detecting acetone, a gas biomarker for diabetes in human exhaled breath, the study can enhance diabetes prediction models and theoretical analysis to a certain extent. Since human exhaled breath contains a large number of compounds, including biomarkers for other diseases, this study holds academic value in non-invasively detecting other components, which is relevant for conditions like pneumonia and chronic airway diseases. Through using self-made sensors, data preprocessing, feature extraction algorithms to obtain higher quality data, and combining them with machine learning algorithms, the study aims to enhance the accuracy of blood sugar prediction models and further advance research on non-invasive blood sugar prediction models [7]. From a practical value perspective: reducing the pain and difficulty of blood or urine collection for diabetes patients, improving patients' quality of life, enhancing patients' compliance with blood sugar testing, enabling multiple repeat tests, and quickly obtaining test results, all contribute to strengthening people's daily health management. Enabling more people to self-monitor blood sugar levels can help slow down the increasing number of diabetes patients. It can also prevent disease transmission risks and environmental pollution that may arise from blood testing, avoiding cross-infection among patients and secondary harm.

Acetone is one of the fatty metabolites in the human body as shown in Fig. 1. The production pathway of acetone in the body is a complex and precise process closely related to energy metabolism, holding significant importance for human health and disease [8].

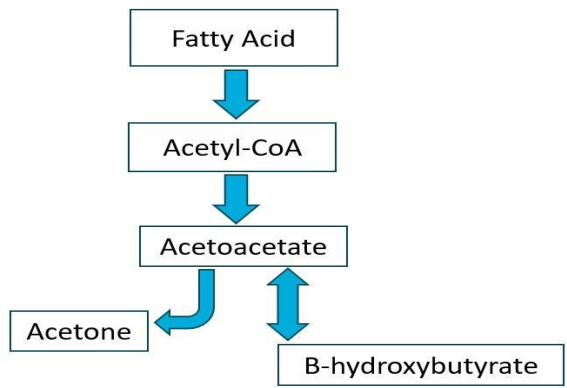

Fig. 1. Acetone metabolism process in the human body

Investigating the production pathway of acetone can enhance our understanding of the body's energy metabolism mechanism, providing a theoretical basis for preventing and treating related diseases. Diabetic patients, due to insufficient insulin for glucose breakdown, rely on breaking down more fats to supply adequate energy for daily needs. This not only leads to excess acetone in the blood but also excessive glucose in the urine [9]. Studies have shown that the average acetone concentration in exhaled breath of adult diabetic patients is higher than 4.7 mg per cubic centimeter, while in normal individuals, acetone content is below 1.9 mg per cubic centimeter. From this perspective, utilizing non-invasive breath testing technology based on acetone sensing for diagnosing diabetes is feasible [10], [11]. Breath testing, as a non-invasive technique, has become a hot research area due to its portability and enhanced patient compliance.

Exhalation detection technology mainly uses gas sensors as the core, combined with related data preprocessing and machine learning algorithms to identify target gases. Gas sensors are essentially transducers that convert changes resulting from their reaction with the target gas into measurable electrical signals. Researchers utilize these electrical signals for further analysis. With the continuous advancement of sensor technology and data analysis algorithms, the development of electronic nose technology combining sensors with pattern recognition has been ongoing [12]. Electronic nose technology, with sensor arrays as the core technology, features broad responses to target gases, ease of operation, low cost, small size, etc., making it applicable in various fields such as early cancer detection, wound infection, respiratory medicine testing, determination of bacterial types in food contamination, odor detection, on-site identification [13], [14], [15], [16], [17], [18], [19], [20], [21].

The earliest research on diabetic patients breathing acetone can be traced back to the early 1950s. Henderson et al. studied the exhaled gases of diabetic and non-diabetic subjects using breath sample pre-concentration techniques and mass spectrometry, focusing on acetone gas in exhaled breath. Pauling et al. discovered over 200 chemical components in human exhaled breath, marking the beginning of modern breath research [22]. Nelson et al. found that compared to exhaled acetone levels in healthy children, diabetic children exhibited higher acetone concentrations. Thakur et al. made an array of six sensors comprised of hybridized graphene oxide (GO) field-effect transistors (FETs) and tested for the detection of acetone. Acetone concentration in a complex VOC mixture was detected with an average 6.65% relative error [23]. Lekshmi et al. used nickel oxide (NiO) as the sensing material for detecting acetone. The simulated values in acetone detection matched well with the actual values, and a function had been established based on the relationship between acetone concentration, sensor resistance, and temperature [24].

Huang et al. used 4 metal oxide semiconductor gas sensors to form an electronic nose and applied it in a flow control system [25]. The results indicated that the backpropagation neural network-based model had better discrimination performance, with a classification accuracy of 82.6%. Associate Professor Wang Fei's research team developed a novel one-dimensional/two-dimensional composite structure nanoarray of $\alpha$-$Fe_2O_3$ and $SnO_2$ capable of detecting trace amounts of acetone, which is valuable for diabetes breath detection [26]. The team led by Liu Huan from Huazhong University of Science and Technology utilized a full-feature odor recognition algorithm to create an intelligent electronic nose [27].

The aim of this study is to build a sensor array by preparing gas sensors and combining them with commercial gas sensors, then integrating data preprocessing and machine learning algorithms to generate a detection model for the diabetes gas biomarker acetone [28]. This model aims to accurately identify acetone in the presence of interfering gases, providing a reference for achieving non-invasive diabetes detection in the future. To meet these goals, the following research was conducted: selecting suitable gas-sensitive materials to prepare acetone gas sensors, studying their sensitivity to acetone, selecting 3 commercial gas sensors and homemade gas sensors based on interfering gases to construct an array, designing circuits, conducting data collection and transmission using an STM32 development board, processing the data collected by the detection device, optimizing potential baseline drift and noise issues in the raw data, and building support vector machine and neural network models. Experimental verification was carried out to assess the discriminative ability and identification accuracy of the recognition model for samples.

## II. EXPERIMENTAL SETUP

### A. Preparation of Acetone Sensors

Acetone sensor is a type of sensor used to detect acetone gas concentration. Its working principle involves detecting the presence of acetone gas by causing a change in resistance through a chemical reaction. When acetone gas passes through the sensor, a chemical reaction occurs inside the sensor, leading to a change in the resistance of the sensitive material. By analyzing the resistance change, the concentration of acetone gas can be determined through concentration-resistance response curves or formulas [29].

Gas sensors face issues such as cross-sensitivity in practical applications, where the detection results from a single sensor may be limited and less reliable. To enhance the accuracy and reliability of detection, it is often necessary to select multiple sensors for sensitivity testing and evaluate the sensors' selectivity to different gases to choose the appropriate gas sensor for specific research. In this study, acetone, a gas biomarker for diabetes, is chosen as the detection target, with methanol and ethanol selected as interfering gases. Three commercial sensors MP502, MQ3, MQ153 and self-prepared sensors are combined to form a sensor array.

Titanium dioxide was selected as the preparation material. Titanium dioxide, an inorganic compound with the chemical formula $TiO_2$, is a white solid or powdery amphoteric compound [30]. Its nanostructure possesses strong oxidizing properties, stability, and good hydrophilicity. The $TiO_2$ nanofilm gas-sensitive sensor has advantages such as simple preparation method, high sensitivity, good consistency, and miniaturization. The wide band gap of $TiO_2$ semiconductor allows for significant improvement in sensor performance by coating and modifying it with other elements. A gas-sensitive sensor for acetone made of $TiO_2$ operates based on the following principle: when pure $TiO_2$ nanomaterial is exposed to air, oxygen molecules combine with electrons in the conduction band of $TiO_2$, forming various valence states of oxygen negative ions on the surface of $TiO_2$ nanomaterial, creating an electron-depleted layer that increases the resistance of the $TiO_2$ gas sensor. When the $TiO_2$ gas sensor is exposed to acetone gas, the adsorbed oxygen negative ions on the surface of $TiO_2$ undergo oxidation-reduction reactions with the target gas. Subsequently, electrons are released back into the semiconductor, increasing the electron concentration, thinning the space charge layer, lowering the barrier potential, and reducing the sensor resistance. Subsequently, based on the sensitive material, device preparation was carried out by initially placing titanium dioxide particles and a small amount of ethanol in a research bowl. Ethanol, due to its rapid evaporation and lack of impact on titanium dioxide, was used as a solvent. The materials were continuously ground in the research bowl using a pestle until a white viscous liquid was formed. This liquid was then spread onto a prepared tube gas sensor dedicated consumable core using a brush. After allowing the material to adsorb, four platinum wires were soldered on both sides of the ceramic tube to the tube gas sensor dedicated consumable base, and a heating resistor was inserted in the middle of the ceramic tube for device heating to the operating temperature. The prepared device underwent signal testing by applying power, adjusting the voltage across the heating resistor to reach the working temperature, generating resistance, confirming the device's proper preparation. Subsequently, the device was placed in an aging chamber for device aging, where the device was kept at the working temperature for 48 hours to ensure device stability.

### B. Gas Sensitivity Test

After aging, the device underwent sensitivity testing to acetone, ethanol, and methanol using a static gas blending method to evaluate stability and sensitivity. To verify if the prepared sensor could be used for detecting the target gas, the optimal operating temperature of the device was determined first. By adjusting the voltage across the heating resistor, the sensor's temperature was increased or decreased, observing the device's response in terms of resistance change. Sensitivity was calculated as the ratio of device resistance in air to device resistance in acetone gas, with higher sensitivity values indicating better performance. The device exhibited the highest sensitivity at 280°C. Subsequently, the stability of the device at this temperature was tested by measuring the response recovery time. The time taken for the device to respond sufficiently to acetone and the time for the device resistance to recover when placed back in air were recorded. A shorter recovery time indicated better device stability. Sensitivity tests for the device with the three gases were conducted, and corresponding concentration curves were plotted.

Experiments were conducted at room temperature with all test samples. The experiments were carried out using a static method, with acetone and varying liquid samples extracted using micro syringes. The working temperature was set at 280°C, and the sensitivity of the device at a 50ppm acetone concentration was tested. During testing, the sensor was placed in a prepared acetone gas bottle for response until the curve stabilized, then removed and placed in an air bottle to observe the sensor recovery before repeating the process in the acetone gas bottle. Finally, the stability curve of the device was obtained as shown in Fig. 2. The first and third stationary curves represent the response curves of the device in air, while the second and fourth stationary curves represent the response curves of the device in acetone gas.

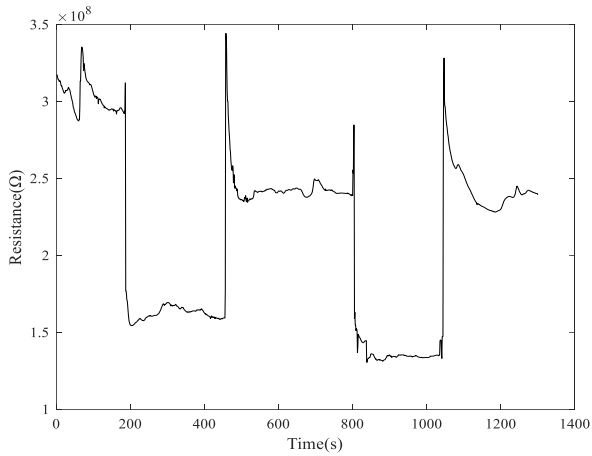

Fig. 2. Sensor Stability Curve

After the above tests, concentration testing is conducted. The response of the device is tested in acetone gas at different concentrations. Firstly, concentration gradients are set up with acetone gas to be tested at 1ppm, 2ppm, 4ppm, 6ppm, 8ppm, 10ppm, 20ppm, 50ppm, 100ppm, 150ppm, 200ppm, 300ppm. During the test, the sensor is placed into the prepared acetone gas bottle, left to respond for 30 seconds, then removed and placed in an air bottle to rest until the sensor recovers before being placed into another gradient acetone gas bottle to repeat the above steps and draw a response curve. Fig. 3 shows the device's response curve to acetone concentration.

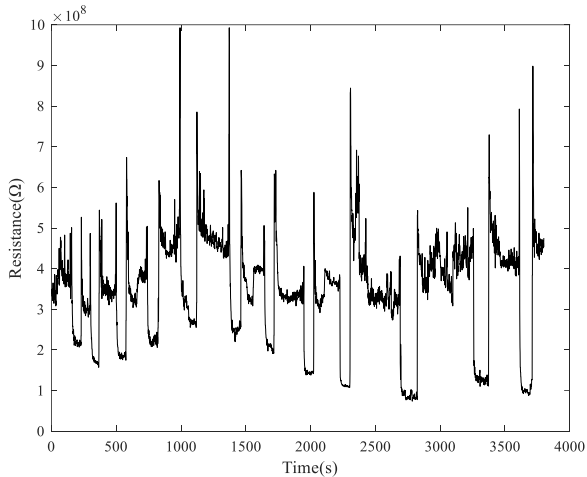

Fig. 3. Sensor concentration curve for acetone

After testing the acetone concentration curve, ethanol and methanol concentration curves are tested. Methanol and ethanol are used as interfering gases for acetone due to their higher content in simulated human breath [31]. Ethanol and methanol gas concentration gradients are set up for testing at 1ppm, 10ppm, 20ppm, 50ppm, 100ppm, 200ppm. During the test, the sensor is placed into the prepared gas bottle to be tested, left to respond for 30 seconds, then removed and placed in an air bottle to rest until the sensor recovers before being placed into another gradient gas bottle to repeat the above steps and draw a response curve. Fig. 4 shows the device's ethanol concentration curve, and Fig. 5 shows the device's methanol concentration curve.

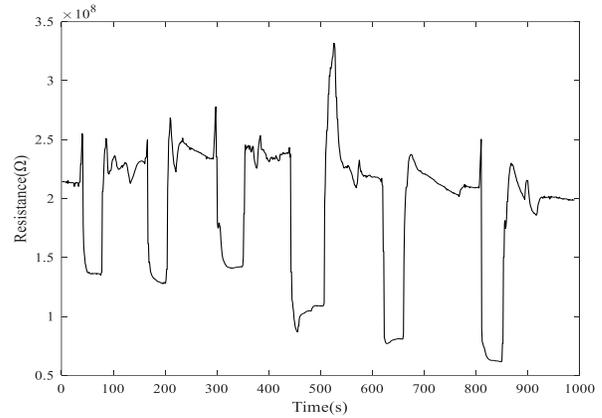

Fig. 4. Sensor concentration curve for ethanol

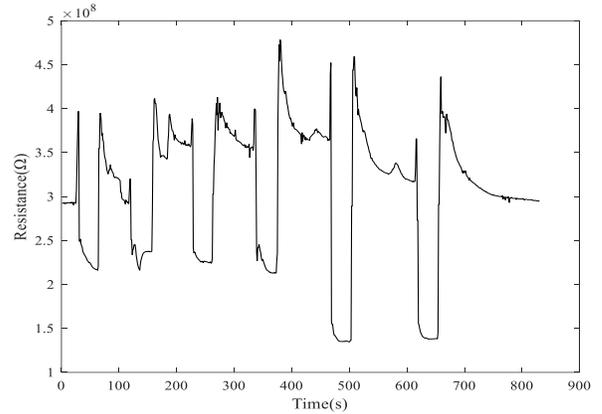

Fig. 5. Sensor concentration curve for methanol

### C. Data Acquisition Method

This study uses an STM32 development board to connect with the sensor and for data transmission between the upper computer and lower computer [32]. The development board is based on the ARM Cortex-M3 core with a 32-bit microcontroller. The MCU is STM32F103RCT6 with a total of 64 pins. The chip's resources include 48KB of static random-access memory, 256KB of flash memory, 2 basic timers, 4 general-purpose timers, 2 advanced timers, 2 DMA with 12 channels, 2 $I^2C$, 3 SPI, a USB interface, a CAN interface, and several IO interfaces. It supports SWD and ISP mainstream debugging methods. The chip system includes a crystal oscillator and reset circuit as shown in Fig. 6.

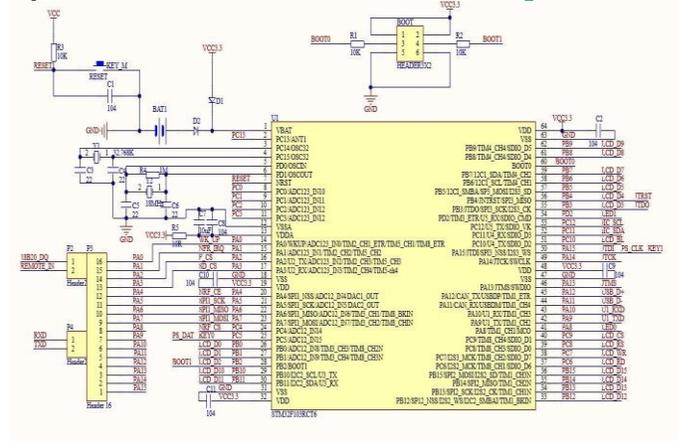

Fig. 6. STM32F103RCT6 chip schematic diagram

The software development tool is Keil uVision5 from the German company LEIL. This software is widely used globally and is currently the best development tool for ARM processors, especially Cortex M core processors. Unlike the 51 single-chip microcontroller, the STM32 has an additional firmware library, which is a collection of functions. The purpose of the firmware library is to handle direct communication with registers below and provide user function call interfaces above. Since the STM32 provides hundreds of registers, it is difficult to remember them all, so the firmware library encapsulates all register-level operations and provides a set of interfaces for developers to call. In many applications, there is no need to operate on specific registers, just knowing which functions to call is sufficient. To master the STM32, it is not enough to just read the firmware library; one must also understand its principles, the operating mechanisms of various peripherals, and have a general understanding of the register configuration process.

When gas passes through the sensor, the sample concentration information is converted into observable electrical signals. Direct current voltage source drives the gas sensor to work, generating a voltage signal that is transmitted to an STM32 for voltage value delivery. To achieve this, the STM32's analog-to-digital converter (ADC) is utilized, which in this model consists of 3 12-bit ADCs and 1 digital-to-analog converter (DAC). The development board can accept a maximum input voltage of 3.3V, so the original ADC sampled value is transformed by *3.3/4096 to obtain the actual input voltage, which is the output voltage of the gas sensor. The ADC in the STM32 is a 12-bit successive approximation type analog-to-digital converter with 18 channels capable of measuring 16 external and 2 internal signal sources. To understand the ADC thoroughly, one needs to examine its related registers including control registers, sampling event registers, regular sequence registers, regular data registers, status registers, etc. In our design, we utilize ADC1 channel 1 for analog-to-digital conversion, where ADC channel 1 is located on PA1. Therefore, enabling the clock for PORTA, setting PA1 as an analog input, configuring PA1 as a multiplexed ADC input, enabling the ADC1 clock, initializing GPIOA1 as an analog input, setting up the ADC, defining the ADC clock division, resolution, mode, scan mode, alignment, etc., is necessary. In the HAL library, the function HAL_ADC_Init is used to initialize the ADC parameters. When all parameters are set, the ADC converter is activated, enabling data retrieval upon input.

For display and communication, a thin film transistor liquid crystal display (TFTLCD) is utilized, capable of displaying 16-bit color true-color images. The process to draw a point on the LCD involves setting coordinates, writing GRAM instructions, inputting color data, and then the corresponding color can be seen on the LCD. The steps for TFTLCD display setup include configuring the IOs connecting the STM32 and TFTLCD module, initializing them, setting up the TFTLCD module, and finally displaying characters and numbers on the TFTLCD module. The information displayed in Fig. 7 includes the voltage value collected by ADC1 and the temperature value and it can monitor the working status of sensors in real time.

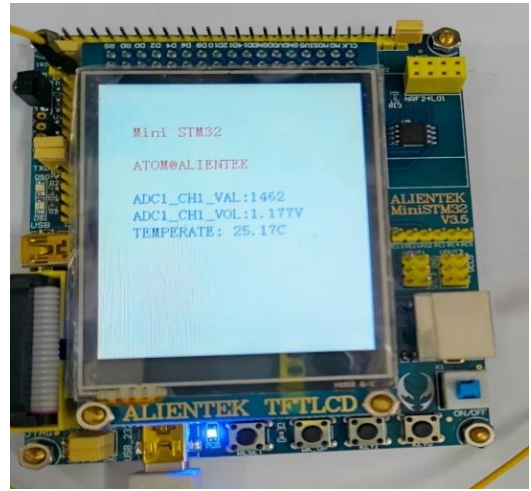

Fig. 7. TFTLCD screen

Serial communication is achieved using the CH340 chip, supporting synchronous and asynchronous transmission on the STM32. Serial port configuration generally involves enabling the serial port clock, GPIO clock, setting pin multiplexer mapping, initializing GPIO with the mode set to multiplexed function, configuring serial port parameters such as baud rate, data length, parity, etc., deciding whether to enable interrupts, enabling the serial port, and writing interrupt handling functions. After setting the parameters, data transmission is achieved through the serial port, which is then transmitted to the computer for saving and waiting for further processing.

## III. Data Analysis Method

The data preprocessing section involves handling missing values, noise filtering, data transformation, etc., on the collected data. In case of missing data due to unexpected interruptions during data collection, methods such as mean interpolation, multiple interpolation, maximum likelihood estimation, etc., can be used to handle the missing values.

During the operation of the detection system, disturbances can affect the sensitivity of the gas sensor, and human testing can introduce errors, leading to fluctuations in the collected data. Each measurement session may experience sensor baseline drift, and when the sensor switches working gas environments, abnormal changes in the measurement curve may occur due to noise. Therefore, data preprocessing is essential before further analysis.

In data preprocessing, polynomial fitting is used to eliminate baseline drift, and uniform moving average filtering is employed to remove noise and reduce data outliers for smoother data. Filtering is done by sliding a fixed-length window M along the signal sequence and calculating the average of the samples within the window to reduce noise. The filtered data can also undergo standardization for uniform processing under the same standard, which can be completed in subsequent operations.

Feature extraction is an essential part of software design, converting sample data into mathematical features for machine learning. The processed data then enters a classification recognition model for further processing. Common methods for feature extraction in this field include scattered feature extraction of raw data, principal component analysis, linear discriminant analysis, and kernel principal component analysis.

This study utilized principal component analysis to analyze preprocessed data with the goal of reducing data dimensions, thereby decreasing computational workload in subsequent classification algorithms [33]. Principal component analysis, also known as principal component analysis technique, is a widely used data dimension reduction method that transforms multiple indicators into a few comprehensive indicators. It can convert high-dimensional data into a lower-dimensional space. The magnitude of variance in the transformed data can serve as a reference standard for how much original data information is contained in each principal component, arranged in order as the first principal component, second principal component, third principal component, and so on. The principle involves identifying the direction with the largest variance in the data, i.e., the principal component, projecting the data onto these principal components to achieve dimension reduction.

PCA is a highly practical data dimensionality reduction method that linearly transforms raw data into a set of variables in a lower-dimensional space, with each variable representing a major component of the original data. Since data output from gas sensors is often nonlinear, and KPCA can handle nonlinear problems by first transforming the data through a kernel function into a new space, then processing it with PCA.

Kernel principal component analysis is a nonlinear extension of principal component analysis suitable for handling nonlinear data, including signals collected by gas sensors in research. The principle involves mapping data to a high-dimensional feature space, using a kernel function to calculate data inner products in the high-dimensional space, thereby achieving nonlinear dimensionality reduction.

Support Vector Machine (SVM) is a type of generalized linear classifier that classifies data through supervised learning, with a decision boundary being the maximum margin hyperplane derived from learning samples. SVM uses hinge loss function to calculate empirical risk and introduces a regularization term in the optimization system to optimize structural risk, exhibiting sparsity and robustness. Through kernel methods, SVM can perform nonlinear classification [34]. SVM's advantages lie in its ability to handle high-dimensional data without being affected by the curse of dimensionality. It has strong generalization capabilities, effectively avoiding overfitting. It is suitable for small sample data, highlighting its ability to handle uneven data distributions. SVM possesses good robustness and interpretability, with the basic principle being as follows: Support Vector Machine is divided into two scenarios: linearly separable and linearly inseparable. In the case of linear separability, there exists a line or a plane that separates the two classes of samples, constructing an optimal classification plane as shown in Fig. 8.

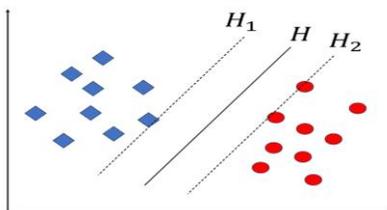

Fig. 8. Classification hyperplane

The figure shows two classes of samples: blue diamonds and red squares, where H represents the classification hyperplane, and H1, H2 are planes passing through H closest to the samples and parallel to the classification hyperplane, with the sample points above referred to as support vectors, and the distance between them known as the classification margin [35]. The optimization goal of SVM is to find a maximum margin hyperplane that maximizes the distance of all sample points to the hyperplane. By introducing slack variables and penalty parameters, SVM can handle cases of linear inseparability, allowing for a certain degree of misclassification while maintaining the largest possible margin boundary.

To achieve precise measurement of the target gas, a backpropagation neural network is employed as the recognition algorithm for acetone, a diabetes gas marker. The topology structure is shown in Fig. 9.

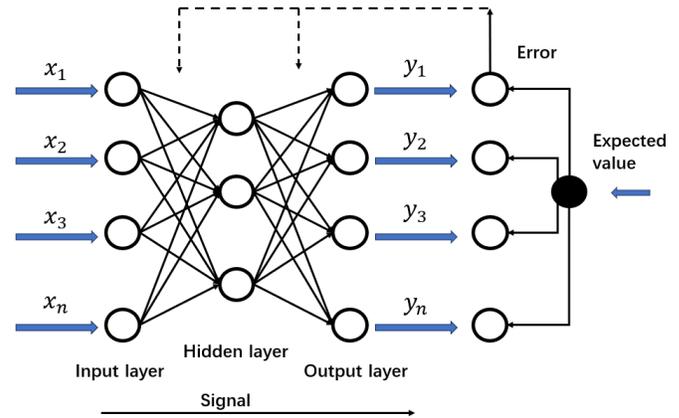

Fig. 9. Topological structure of BP neural network

The topology of the backpropagation neural network can be seen in the figure, consisting of the input layer, hidden layer, and output layer, with no connections between modules within each layer. Initially, after data selection, signals propagate forward to the neurons in the hidden layer. Neurons sum up these signals to obtain a total input value, which is then compared with the neuron's threshold, processed through a function to produce an output. This output then serves as the input for the subsequent layers of neurons, propagating layer by layer. Its main feature is the forward propagation of signals and backward propagation of errors. The process of modifying weights through forward and backward propagation is the learning training process of the neural network. The learning training process of the neural network can continue in a loop until reaching a predetermined error precision or specified number of training iterations [36].

The BP neural network mainly consists of two processes: forward propagation to obtain the desired output and backward propagation to reduce the error. Establishing the network model requires training on samples, where the quality of the samples determines the model's recognition accuracy. Once the network model is constructed, predictions can be made, with each prediction result varying.

The number of neurons in the input layer of the neural network is determined by the original collected data, primarily used for acquiring input information. The output layer represents the predicted values. The number of hidden layers in the actual program needs to be determined through repeated debugging. This layer primarily adjusts weights to make neurons react to certain patterns.

## IV. Detection of Acetone

The functional verification of the detection system involves the detection of the diabetes gas marker acetone. Utilizing the designed detection system, the selectivity of acetone was studied in the presence of air and interfering VOC mixtures. Different test gases were prepared by changing the proportion of acetone. Experiments were conducted at room temperature with all test samples present. The experiments were carried out using a static method, utilizing acetone, ethanol, and methanol, with varying amounts extracted using micro-syringes.

Acetone, ethanol, and methanol were studied, configuring acetone and ethanol binary gas mixtures, as well as acetone and methanol binary gas mixtures. In each test, the sensor was left to stabilize for 30 seconds in the test gas cylinder, then quickly removed and placed in the air cylinder until the sensor recovered before repeating the experiment.

The mixing ratios in the ethanol-acetone binary gas experiment are shown in Table II. 600 data sets were selected as training samples, with 80 data sets as test samples, acetone labeled as 1, and ethanol as 2.

TABLE II
ACETONE ETHANOL GAS MIXTURE CONCENTRATION RATIO

| Acetone | Ethanol |
|---|---|
| 100ppm | 0ppm |
| 99ppm | 1ppm |
| 90ppm | 10ppm |
| 50ppm | 50ppm |
| 50ppm | 0ppm |
| 49.5ppm | 0.5ppm |
| 45ppm | 5ppm |
| 25ppm | 25ppm |

The collected data on the mixed gases were processed. Initially, a filtering process was applied as shown in Fig. 10 to remove any abnormal data generated during dynamic experiments.

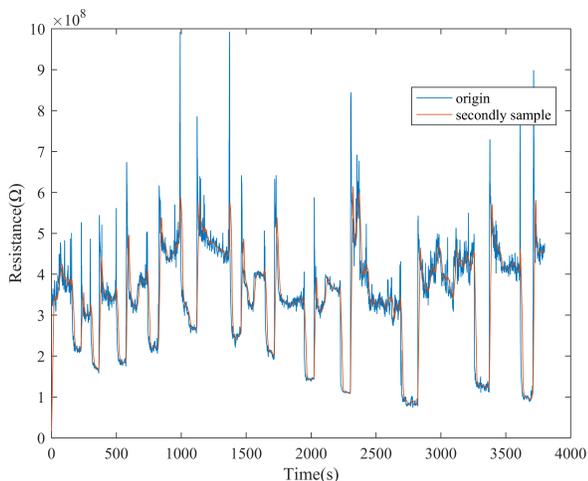

Fig. 10. Filter comparison chart

After preprocessing the raw data, they were input into a support vector machine for classification, resulting in a gas recognition rate of 96%. The classifier's classification results are shown in Fig. 11.

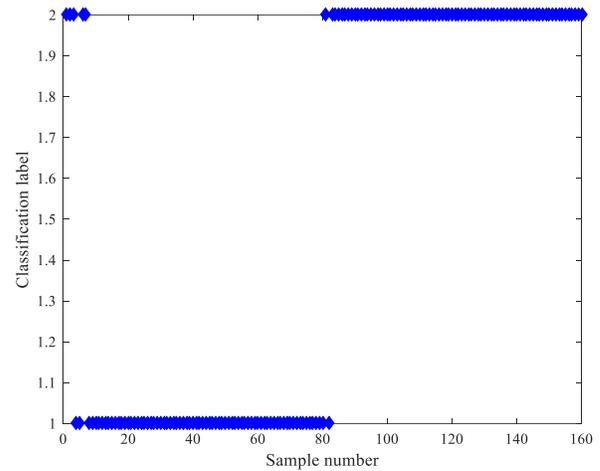

Fig. 11. Classification results of acetone ethanol gas

The mixing ratios in the methanol-acetone binary gas experiment are shown in Table III. 700 data sets were selected as training samples, with 100 data sets as test samples, acetone labeled as 1, and methanol as 3. The collected data were filtered.

TABLE III
ACETONE METHANOL GAS MIXTURE CONCENTRATION RATIO

| Acetone | Methanol |
|---|---|
| 100ppm | 0ppm |
| 99ppm | 1ppm |
| 90ppm | 10ppm |
| 50ppm | 50ppm |
| 50ppm | 0ppm |
| 49.5ppm | 0.5ppm |
| 45ppm | 5ppm |
| 25ppm | 25ppm |

After preprocessing the raw data, they were input into a support vector machine for classification, resulting in a gas recognition rate of 97.5%. The classifier's classification results are shown in Fig. 12.

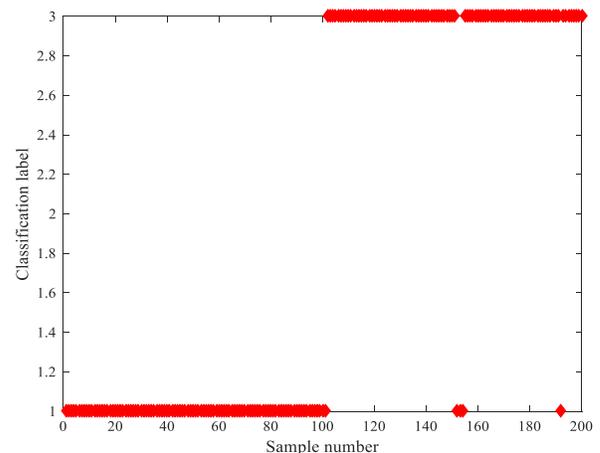

Fig. 12. Classification results of acetone methanol gas

Acetone, ethanol, and methanol were studied, configuring acetone, ethanol, and methanol ternary gas mixtures. In each test, the sensor was left to stabilize for 30 seconds in the test gas cylinder, then quickly removed and placed in the air cylinder until the sensor recovered before repeating the experiment.

The mixing ratios in the qualitative measurement experiment of the ternary gas mixture are shown in Table IV. 550 data sets were selected as training samples, with 50 data sets as test samples, acetone labeled as 1, ethanol as 2, and methanol as 3. The collected data were first filtered to remove any abnormal data generated during dynamic experiments.

TABLE IV
ACETONE ETHANOL AND METHANOL GAS MIXTURE CONCENTRATION RATIO

| Acetone | Ethanol | Methanol |
|---|---|---|
| 200ppm | 0ppm | 0ppm |
| 198ppm | 1ppm | 1ppm |
| 180ppm | 10ppm | 10ppm |
| 100ppm | 50ppm | 50ppm |
| 100ppm | 0ppm | 0ppm |
| 98ppm | 0.5ppm | 0.5ppm |
| 90ppm | 5ppm | 5ppm |
| 50ppm | 25ppm | 25ppm |

After preprocessing the raw data, they were input into a support vector machine for classification, the predicted results by the classifier are shown in Fig. 13, resulting in a gas recognition rate of 90%.

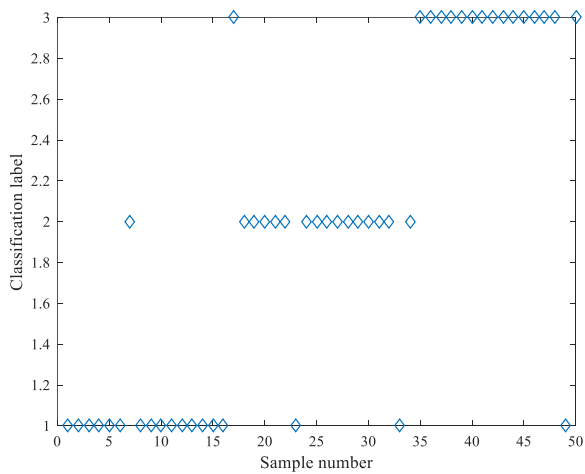

Fig. 13. Classification results of acetone ethanol and methanol gas

## V. CONCLUSION

This study focuses on the diabetic gas biomarker acetone, and a detection system based on gas sensors has been designed to achieve ppm-level classification and concentration prediction. The detection device consists of a sensor array and a microcontroller, with an acetone sensor prepared using titanium dioxide. The microcontroller communicates with the host computer via serial communication, and it is also equipped with a touchscreen module and a 3.3V power module. The raw data undergo baseline drift and smoothing processing, followed by feature processing using kernel principal component analysis. The processed data is then input into the model for training, and once the model is trained, it is tested on a test set in proportion. The results show that this detection system achieves recognition accuracy of over 90% for mixed gases, enabling classification and preliminary detection of acetone. However, due to large data gradients, there is significant error in predicting acetone concentration, necessitating further research.

## REFERENCES

[1] A. Boulton, "Strengthening the International Diabetes Federation(IDF)," *Diabetes Res. Clin. Pract.*, vol. 160, Feb. 2020, Art. no. 108029.
[2] A. Peña et al., "Real-time monitoring of breath biomarkers with a magnetoelastic contactless gas sensor: a proof of concept," *Biosensors*, vol. 12, no. 10, Oct. 2022, Art. no. 871.
[3] F. A. Hashim, E. H. Houssein, K. Hussain, M. S. Mabrouk and W. Al-Atabany, "Honey badger algorithm: new metaheuristic algorithm for solving optimization problems," *Math. Comput. Simul.*, vol. 192, pp. 84-110, Feb. 2022.
[4] F. A. Hashim, K. Hussain, E. H. Houssein, M. S. Mabrouk and W. Al-Atabany, "Archimedes optimization algorithm: a new metaheuristic algorithm for solving optimization problems," *Appl. Intell.*, vol. 51, no. 3, pp. 1531-1551, Mar. 2021,
[5] N. Bhaskar, V. Bairagi, E. Boonchieng and M. V. Munot, "Automated detection of diabetes from exhaled human breath using deep hybrid architecture," *IEEE Access*, vol. 11, pp. 51712-51722, Jul. 2023.
[6] C. L. Xue et al., "Smartphone case-based gas sensing platform for on-site acetone tracking," *ACS Sens.*, vol. 7, no. 5, pp. 1581-1592, May. 2022.
[7] A. R. Cingireddy, R. Ghosh, V. K. Melapu, S. Joginipelli and T. A. Kwembe, "Classification of parkinson's disease using motor and non-motor biomarkers through machine learning techniques," *IJQSPR.*, vol. 15, no. 1, pp. 1-21, Apr. 2022.
[8] G. Hancock et al., "The correlation between breath acetone and blood betahydroxybutyrate in individuals with type 1 diabetes," *J. Breath Res.*, vol. 13, no. 1, Jan. 2021, Art. no. 017101.
[9] J. Sorocki and A. Rydosz, "A prototype of a portable gas analyzer for exhaled acetone detection," *Appl. Sci. Basel*, vol. 9, no. 13, Jul. 2019, Art. no. 2605.
[10] K. D. Wu, Y. F. Luo, Y. Li and C. Zhang, "Synthesis and acetone sensing properties of $ZnFe_2O_4$/rGO gas sensors," *Beilstein J. Nanotechnol.*, vol. 10, pp. 2516-2526, Dec. 2019.
[11] P. DAS, B. C. Behera, S. K. Panigrahy, A. K. Sahu and S. K. Tripathy, "Enhanced acetone sensing at room temperature using tailored $Co_3O_4$ nanostructures-coated optical fibers: an application towards non-invasive blood glucose sensor," *Surf.*, vol. 41, Oct. 2023, Art. no. 103256.
[12] X. H. Weng et al., "A preliminary screening system for diabetes based on in-car electronic nose," *Endocr. Connect.*, vol. 12, no. 3, Mar. 2023, Art. no. e220437.
[13] Z. Y. Ye, J. Wang, H. Hua, X. D. Zhou and Q. L. Li, "Precise detection and quantitative prediction of blood glucose level with an electronic nose system," *IEEE Sens. J.*, vol. 22, no. 13, pp. 12452-12459, Jul. 2022.
[14] I. Oakley-Girvan and S. W. Davis, "Breath based volatile organic compounds in the detection of breast, lung, and colorectal cancers: a systematic review," *Cancer Biomark.*, vol. 21, no. 1, pp. 29-39, Jan. 2018.
[15] C. H. Huang et al., "A study of diagnostic accuracy using a chemical sensor array and a machine learning technique to detect lung cancer," *Sensors*, vol. 18, no. 9, Sep. 2018, Art. no. 2845.
[16] H. X. Luo, P. F. Jia, S. Q. Qiao and S. K. Duan, "Enhancing electronic nose performance based on a novel QPSO-RBM technique," *Sens. Actuators B, Chem.*, vol. 259, pp. 241-249, Apr. 2018.
[17] R. Gasparri, G. Sedda and L. Spaggiari, "The electronic nose's emerging role in respiratory medicine," *Sensors*, vol. 18, no. 9, Sep. 2018, Art. no. 3029.
[18] M. Ezhilan, N. Nesakumar, K. J. Babu, C. S. Srinandan and J. B. B. Rayappan, "A multiple approach combined with portable electronic nose for assessment of post-harvest sapota contamination by foodborne pathogens," *Food Bioprocess Tech.*, vol. 13, no. 7, pp. 1193-1205, Jul. 2020.
[19] M. Aleksic, A. Simeon, D. Vujic, S. Giannoukos and B. Brkic, "Food and lifestyle impact on breath VOCs using portable mass spectrometer—pilot

study across European countries," *J. Breath Res*., vol. 17, no. 4, Oct. 2023, Art. no. 046004.

[20] G. Lagod, S. M. Duda, D. Majerek, A. Szutt and A. Dolhanczuk-Srodka, "Application of electronic nose for evaluation of wastewater treatment process effects at full-scale WWTP," *Processes*, vol. 7, no. 5, May. 2019, Art. no. 251.

[21] A. Blanco-Rodriguez, "Development of an electronic nose to characterize odours emitted from different stages in a wastewater treatment plant," *Water Res*., vol. 134, pp. 92-100, May. 2018.

[22] Z. N. Wang and C. J. Wang, "Is breath acetone a biomarker of diabetes? A historical review on breath acetone measurements," *J. Breath. Res*., vol. 7, no. 3, Aug. 2013, Art. no. 037109.

[23] U. N. Thakur, R. Bhardwaj and A. Hazra, "A multivariate computational approach with hybrid graphene oxide sensor array for partial fulfillment of breath acetone sensing," *IEEE Sens. J.*, vol. 22, no. 21, pp. 20207-20215, Nov. 2022.

[24] M. S. Lekshmi, K. Arun and K. J. Suja, "A microcontroller-based signal conditioning circuitry for acetone concentration detection using a metal oxide-based gas sensor," *J. Comput. Electron.*, vol. 21, no. 4, pp. 1017-1025, Aug. 2022.

[25] J. M. Huang and J. Wu, "Robust and rapid detection of mixed volatile organic compounds in flow through air by a low cost electronic nose," *Chemosensors*, vol. 8, no. 3, Sep. 2020, Art. no. 73.

[26] H. M. Gong, C. H. Zhao, G. Q. Niu, W. Zhang and F. Wang, "Construction of 1D/2D α-$Fe_2O_3$/$SnO_2$ hybrid nanoarrays for sub-ppm acetone detection," *Research*, vol. 2020, Feb. 2020, Art. no. 2196063.

[27] C. Fang et al., "Smart electronic nose enabled by an all-feature olfactory algorithm," *Adv. Intell. Syst.*, vol. 4, no. 7, Jul. 2022, Art. no. 2270032.

[28] L. B. Cai et al., "Ultrasensitive acetone gas sensor can distinguish the diabetic state of people and its high performance analysis by first-principles calculation," *Sens. Actuators B, Chem.*, vol. 351, Jan. 2022, Art. no. 130863.

[29] G. L. Li et al., "Mid-infrared acetone sensor for exhaled gas using FWA-LSSVM and empirical mode decomposition algorithm," *Measurement*, vol. 2013, May. 2023, Art. no. 112716.

[30] T. Gakhar and A. Hazra, "Oxygen vacancy modulation of titania nanotubes by cathodic polarization and chemical reduction routes for efficient detection of volatile organic compounds," *Nanoscale*, vol. 12, no. 16, pp. 9082-9093, Apr. 2020.

[31] H. Y. Zhu, C. Liu, Y. Zheng, J. Zhao and L. Li, "A hybrid machine learning algorithm for detection of simulated expiratory markers of diabetic patients based on gas sensor array," *IEEE Sens. J.*, vol. 23, no. 3, pp. 2940-2947, Feb. 2023.

[32] K. Park et al., "An energy-efficient multimode multichannel gas-sensor system with learning-based optimization and self-calibration schemes," *IEEE Trans. Ind. Electron*., vol. 67, no. 3, pp. 2402-2410, Mar. 2020.

[33] A. Gade, V. Vijayabaskar and J. Panneerselvam, "A non-invasive blood glucose monitoring for diabetics with breath biomarkers: an ensemble-of-classifiers model," *J. Mech. Med. Biol.*, vol. 23, no. 10, Dec. 2023, Art. no. 2350008.

[34] H. B. Wu, S. L. Shi, Y. Lu, Y. Liu and W. H. Hung, "Top corner gas concentration prediction using t-distributed stochastic neighbor embedding and support vector regression algorithms," *Concurr. Comp-Pract. E.*, vol. 32, no. 14, Jul. 2020, Art. no. e5705.

[35] J. Cervantes, F. Garcia-Lamont, L. Rodriguez-Mazahua and A. Lopez, "A comprehensive survey on support vector machine classification: applications, challenges and trends," *Neurocomputing*, vol. 408, pp. 189-215, Sep. 2020.

[36] N. Bhaskar and M. Suchetha, "A computationally efficient correlational neural network for automated prediction of chronic kidney disease," *IRBM*, vol. 42, no. 4, pp. 268-276, Aug. 2021.